# The orbital and superhump periods of the deeply eclipsing dwarf nova SDSS J150240.98+333423.9

Jeremy Shears, Tut Campbell, Jerry Foote, Russ Garrett, Tim Hager, William Mack Julian, Jonathan Kemp, Gianluca Masi, Ian Miller, Joseph Patterson, Michael Richmond, Frederick Ringwald, George Roberts, Javier Ruiz, Richard Sabo, William Stein

## Abstract

During July 2009 we observed the first confirmed superoutburst of the eclipsing dwarf nova SDSS J150240.98+333423.9 using CCD photometry. The outburst amplitude was at least 3.9 magnitudes and it lasted at least 16 days. Superhumps having up to 0.35 peak-to-peak amplitude were present during the outburst, thereby establishing it to be a member of the SU UMa family. The mean superhump period during the first 4 days of the outburst was $P_{sh}$ = 0.06028(19) d, although it increased during the outburst with $dP_{sh}/dt$ = + 2.8(1.0) x $10^{-4}$. The orbital period was measured as $P_{orb}$ = 0.05890946(5) d from times of eclipses measured during outburst and quiescence. Based on the mean superhump period, the superhump period excess was ε = 0.023(3). The FWHM eclipse duration declined from a maximum of 10.5 min at the peak of the outburst to 3.5 min later in the outburst. The eclipse depth increased from ~0.9 mag to 2.1 mag over the same period. Eclipses in quiescence were 2.7 min in duration and 2.8 mag deep.

## Introduction

SDSS J150240.98+333423.9 (hereafter "SDSS 1502") was first identified spectroscopically as a dwarf nova in the Sloan Digital Sky Survey (SDSS) database [1]. A deep doubling of its Balmer emission lines suggested that it was a high inclination system with the likelihood of eclipses. Follow-up photometry confirmed the presence of 2.5 mag deep eclipses [1]. Subsequently the orbital period was measured as 0.05890961(5) d (~84.8 min) and the mass ratio, q, of the secondary to the white dwarf primary was determined as 0.109, with the white dwarf having a mass of about 0.82 solar masses [2].

The orbital period places SDSS 1502 well below the so-called period gap in the distribution of orbital periods of dwarf novae, suggesting it was likely to be a member of the SU UMa family.

## Outbursts of SDSS 1502

Figure 1 shows the light curve of SDSS 1502 between 2005 Mar 15 and 2009 Oct 10. Much of the data are from the Catalina Real-Time Transient Survey (CRTS) [3], supplemented with data from the authors. This shows that at quiescence the star varied between mag 17 and 17.5 with a mean of 17.2, although clearly there are excursions to fainter magnitudes which presumably coincide with eclipses. At least two outbursts are apparent: one detected by CRTS in March 2007 reaching mag 14.1 and a further one in July 2009 reaching mag 13.3. Time resolved photometry of the latter outburst is presented in this paper. It is interesting to note that CRTS did not cover this field during the July 2009 outburst, there being an observational gap between 2009 May 27 and July 31. Two further brightening events reaching mag 16.0 and 16.1 were detected, which may represent further outbursts. Other outbursts might have been missed due to incomplete observational coverage, so the 853 day interval between the two main outbursts may be much longer than the actual outburst period.

**Photometry and analysis**

The authors conducted photometry using the instrumentation shown in Table 1 and according to the observation log in Table 2. Most of the data from the July 2009 outburst (Table 2(a)) were obtained at the observing stations of the Center for Backyard Astrophysics (CBA), a world-wide network of small telescopes. Data on eclipses at quiescence were obtained with the 1.3 m MDM telescope on Kitt Peak (Table 2(b)). Images were dark-subtracted and flat-fielded prior to being measured using differential aperture photometry relative to the sequence in AAVSO chart 1432hli [4]. Heliocentric corrections were applied to all data.

**The July 2009 outburst**

The outburst of SDSS 1502 discussed in this paper was detected by JS on 2009 July 9.949 at 13.7C (C, unfiltered CCD) [5]. The overall light curve of the outburst is shown in the top panel of Figure 2. The apparent scatter in the magnitudes is of course mainly due to presence of the eclipses. SDSS 1502 was observed to be at its brightest on discovery night at 13.3C, representing an amplitude of 3.9 magnitudes above mean quiescence. Sixteen days later the star was almost back at quiescence; the beginning of the outburst is not well constrained since the observation prior to the detection was on July 3.942 (>16.7C), some 6 days earlier. Considering the average magnitude outside eclipses, the brightness showed an approximately linear decline at 0.25 mag/d, although there is some evidence that there was a steeper decline near the beginning of the outburst (JD 2455022 to 2455025), followed by a plateau (JD 2455026 to 24550228) then a faster decline towards quiescence.

In Figure 3 we plot expanded views of the time series photometry, having subtracted the linear trend, where each panel shows one day of data drawn to the same scale. This clearly shows recurrent eclipses superimposed on an underlying superhump modulation, each of which will be considered in more detail later. The presence of superhumps is diagnostic that SDSS 1502 is a member of the SU UMa family of dwarf novae making this the first confirmed superoutburst of the star.

**Measurement of the orbital period**

Times of minimum were measured for eclipses observed during the outburst using the 5-term Fourier fit in the *Minima* software package [6]. We supplemented these with times of minimum measured from quiescence photometry obtained using the

1.3 m MDM telescope on Kitt Peak. Table 3 lists the times of minimum, where the 104 observed eclipses are labelled with the corresponding orbit number starting from 0. The orbital period was then calculated from a linear fit to these times of minima as $P_{orb}$ = 0.05890946(5) d. The eclipse time of minimum ephemeris is:

$$HJD_{min} = 2453849.94908135(2) + 0.05890946(5) \times E \quad \text{Equation 1}$$

The O-C (Observed – Calculated) residuals of the eclipse minima relative to this ephemeris are shown in Figure 4. This suggests that the period remained constant during the period of observations. However, we note that there is considerable scatter in the O-C values which were measured during the 2009 outburst of up to 0.02 cycles (~1.7 min). This is due to the difficulty in isolating eclipse minima relative to other large-scale changes in the light curve in the form of the underlying superhumps and the fact that each eclipse was defined by rather few data points.

Our value of $P_{orb}$ is consistent at the 2-sigma level with the one reported in reference 2, which was derived from observations obtained over a shorter period of time.

**Measurement of the superhump period**

Analysis of the superhumps was complicated by the presence of the eclipses, which often distorted the shape of the superhumps. The peak-to-peak superhump amplitude was ~0.35 mag on the first night (JD 2455022), declining to ~0.2 mag on 2455026 and subsequently increasing again gradually until the final night of time series photometry (JD 2455038) when the amplitude was 0.2 mag.

To study the superhump behaviour, we first extracted the times of each sufficiently well-defined superhump maximum by fitting a quadratic function to the individual light curves. We omitted superhumps whose maxima coincided with eclipses. Times of 37 superhump maxima were found and are listed in Table 4. An analysis of the times of maximum for cycles 0 to 55 (JD 2455022 to 2455026) allowed us to obtain the following linear superhump maximum ephemeris:

$$HJD_{max} = 2455022.50822(24) + 0.06028(19) \times E \quad \text{Equation 2}$$

This gives the mean superhump period for the first four days of the superoutburst $P_{sh}$ = 0.06028(19) d. The O–C residuals for the superhump maxima for the complete outburst relative to the ephemeris are shown in the bottom panel in Figure 2. There is a suggestion that the superhump period remained constant during the first 4 days of the outburst and then gradually increased through the rest of the outburst. The data are also consistent with a period increase during the outburst with $dP_{sh}/dt$ = + 2.8(1.0) x $10^{-4}$.

To confirm our measurements of $P_{sh}$, we carried out a period analysis of the data using the Date Compensated Discrete Fourier Transform (DCDFT) algorithm in Peranso [7]. To investigate the stability of the period, we divided the light curve into three sections and analysed each section separately. Figure 5 shows the power spectrum of the data from JD 2455022 to 2455025, which has its highest peaks at a period of 0.050891(13) d and 0.06024(15) d. The period error estimate is derived using the Schwarzenberg-Czerny method [8]. We interpret the shorter period signal as $P_{orb}$ and the longer one as $P_{sh}$. Both values are consistent with our earlier measurements from the times of eclipse minima and superhump maxima. Removing $P_{orb}$ by pre-whitening gave the power spectrum in Figure 6. In this case the strongest signal was at 0.06024(15) d (plus its 1, 2 and 3 c/d aliases), which had also been present in the original power spectrum as the strongest peak, and which corresponds to $P_{sh}$. We performed a similar analysis for the remaining sections of the light curve, and for the combined outburst data, with the following results:

JD 2455022 to 2455025    $P_{sh}$ = 0.06024(15) d

JD 2455026 to 2455029    $P_{sh}$ = 0.06029(12) d

JD 2455030 to 2455038    $P_{sh}$ = 0.06098(28) d

Combined data            $P_{sh}$ = 0.06042(9) d

These results are also consistent with an increase in $P_{sh}$ during the outburst.

**The nature of the eclipses**

One of the most interesting aspects of SDSS 1502 is its deep eclipses. We measured the eclipse duration as the full width at half minimum (FWHM; Table 3). Figure 7 shows that the eclipse duration was greatest at the peak of the outburst (10.5 min) and declined as the outburst progressed, with the final eclipses being about one-third the duration (3.5 min) of the first ones observed. This is a common feature of eclipses during dwarf nova outbursts and is due to the accretion disc being largest at the peak of the outburst and subsequently shrinking from the outside inwards as material drains from the disc as the outburst progresses [9]. We measured the eclipse duration at quiescence, when the accretion disc diameter is expected to be at a minimum, as 2.7 min (average of 5 quiescence eclipses on 2006 Apr 25).

Figure 8 shows that there was also a trend of increasing eclipse depth during the outburst from ~0.9 mag near the beginning to ~2.1 mag towards the end (data are given in Table 3). During quiescence the eclipses were ~2.8 magnitudes deep. A cursory examination of the time series light curves presented in Figure 3 shows that the eclipse depth is also affected by the location of the superhump: in general eclipses are shallower when hump maximum coincides with eclipse. This is commonly observed in eclipsing SU UMa systems including DV UMa, IY UMa and SDSS J122740.82+513925.0 [9, 10, 11] where there is a relationship between eclipse depth and the precession period, $P_{prec}$ (i.e. the beat period of the superhump and orbital periods). However, we could find no such correlation in SDSS 1502, presumably because any variation in depth associated with the beat period would have been masked by the much larger increase in depth as the star faded.

According to the relation $1/P_{prec} = 1/P_{orb} - 1/P_{sh}$, the precession period should be about 2.6 d, based on our measured values of $P_{sh}$ and $P_{orb}$. However, we searched for such a signal corresponding to the precession period in the DCDFT power spectrum in the interval 1 to 5 d without success.

A typical example of a quiescence light curve is shown in Figure 9 where 6 eclipses 2.8 magnitude deep are present. Another prominent feature is an orbital hump which occurs just before each eclipse. Orbital humps are due to the presence of a "bright spot" which forms where the material flowing from the secondary star hits the edge of the accretion disc [12]. To study the eclipse profile in more detail, we took the flux intensity data from Figure 9 (rather than the magnitude data, as is standard practise when investigating deep eclipses) and folded it on $P_{orb}$. This gave the average eclipse profile shown in Figure 10, where we plot the mean value of bins containing five separate intensity measurements. The eclipse was not quite symmetrical. Littlefair *et al*. [2] also found an asymmetric eclipse in SDSS 1502 and they used multicolour eclipse mapping and modelling to associate different parts of the eclipse profile with different parts of the system undergoing eclipse, including the white dwarf, the bright spot and the accretion disc. Unfortunately, the resolution of our data was not sufficient to identify with certainty the different parts of the system undergoing eclipse, although the rapid rise in brightness at the end of the eclipse, at a phase of 0.2, may due to the white dwarf coming out of eclipse.

**Discussion**

Taking our mean superhump period for the first four days of the superoutburst, $P_{sh}$ = 0.06028(19) d, and our measured orbital period, $P_{orb}$ = 0.05890946(5) d, we calculate the superhump period excess ε = 0.023(3). Such value is consistent with other SU UMa systems of similar orbital period [12].

Patterson *et al*. [13] established an empirical relationship between ε and q, the secondary to primary mass ratio: ε = 0.18*q + 0.29*$q^2$. This assumes a white dwarf of ~0.75 solar masses which is typical for SU UMa systems, but which is slightly smaller than the white dwarf in SDSS 1502. Taking q = 0.109 for SDSS 1502, this empirical relationship predicts ε = 0.023. Such value is the same as our measured value of ε, again confirming that SDSS 1502 is a typical SU UMa system.

Kato *et al*. [14] have studied the superhump period changes in a large numbers of SU UMa systems and in general find three distinct stages: an early evolutionary stage (A) with longer superhump period, a middle stage (B) for a large part of the outburst in which systems with $P_{orb}$ < 0.08 d have a positive period derivative, and a final stage (C) with a shorter $P_{sh}$. In the case of the outburst of SDSS 1502, it is possible that we missed stage A. The bulk of our observations were probably from stage B and our positive superhump period is consistent with *Kato et al.*'s observations. Stage C usually occurs towards the end of the outburst, during which we were unable to measure the superhump period since the superhumps occurred too close to the eclipses during this phase. Although we reported a continuous period increase during the outburst ($dP_{sh}/dt = + 2.8(1.0) \times 10^{-4}$), close inspection of the trend in the O-C data in Figure 2 suggests that the situation may actually be more complex. From the beginning of the outburst to about JD 24455028 the period is generally increasing, but there then appears to be a period decrease between JD 24455028 and JD 24455030, following which it increases again. By comparing the O-C diagram with the light curve it is tempting to speculate that the changes in O-C may correspond to the inflexions in the light curve (such as the plateau between JD 2455026 to 24550228) mentioned in the section on the *July 2009 Outburst*. These inflexions may indicate that the accretion disc is not shrinking at a constant rate during the outburst, which may in turn affect the precession of the disc and hence the form of the O-C curve. The plot in Figure 7 suggests that there may be a discontinuity in the eclipse duration data between JD 2455028 and 2455030 which in turn could indicate a change in the rate of contraction of the disc. A similar possible discontinuity at about the same time can also be seen on the eclipse depth plot (Figure 8). Whilst these observations are intriguing, a link between them and the physical state of the accretion disc must remain speculation. Observations of future outbursts may reveal further information.

**Conclusions**

Analysis of the first confirmed superoutburst of SDSS 1502 during July 2009 has shown that it is a member of the SU UMa family of dwarf novae. The outburst amplitude was at least 3.9 magnitudes and it lasted at least 16 days. Analysis of eclipse times from outburst and quiescence yielded an orbital period of $P_{orb} = 0.05890946(5)$ d. Time-series photometry during the outburst revealed superhumps with a maximum peak-to-peak amplitude of 0.35 magnitudes. The mean superhump period during the first 4 days of the outburst was $P_{sh} = 0.06028(19)$ d, although the superhump period increased during the outburst with $dP_{sh}/dt = + 2.8(1.0) \times 10^{-4}$. Based on the mean superhump period, the superhump period excess was $\varepsilon = 0.023(3)$. The FWHM eclipse duration declined from a maximum of 10.5 min at the peak of the outburst to 3.5 min later in the outburst, indicating a shrinking accretion disc. The depth of the eclipses increased from ~0.9 mag near the beginning of the outburst to 2.1 mag at the end. Eclipses in quiescence were 2.7 min in duration and 2.8 mag deep. Long term monitoring of the star between March 2005 and Oct 2009 has revealed at least one, and possibly three, additional outbursts.

**Acknowledgements**

The authors gratefully acknowledge the use of data from the Catalina Real-Time Transient Survey, kindly provided by Dr. Andrew Drake. We also used SIMBAD, operated through the Centre de Donées Astronomiques (Strasbourg, France), and the NASA/Smithsonian Astrophysics Data System. FR acknowledges the use of the Fresno State's observing station at the Sierra Remote Observatories and JR thanks the Government of Cantabria for access to the Observatory of Cantabria. JS thanks the Council of the British Astronomical Association for the award of a Ridley Grant that was used to purchase equipment that was used in this research. Finally we thank the referees whose comments have helped to improve the paper.

**Addresses**

JS: "Pemberton", School Lane, Bunbury, Tarporley, Cheshire, CW6 9NR, UK [bunburyobservatory@hotmail.com]
TC: 7021 Whispering Pine, Harrison, AR 72601, USA [jmontecamp@yahoo.com]
JF:  Center for Backyard Astrophysics (Utah), 4175 E. Red Cliffs Drive, Kanab, UT 84741, USA [jfoote@scopecraft.com]
RG: 4714 Innsbrooke Pkwy, Pinson,  Alabama 35126, USA [russ@russgarrett.com]
TH: 34 Mount Tom Rd., New Milford, CT 06776, USA [thager6164@earthlink.net]
WMJ: 4587 Rockaway Loop, Rio Rancho, NM 871224, USA [mack-julian@cableone.net]
JK: Joint Astronomy Centre, University Park, 660 North A'ohōkū Place, Hilo, HI 96720, USA [j.kemp@jach.hawaii.edu]
GM: Center for Backyard Astrophysics (Italy), via Madonna de Loco 47, 03023 Ceccano FR, Italy  [gianluca@bellatrixobservatory.org]
IM: Furzehill House, Ilston, Swansea, SA2 7LE, UK [furzehillobservatory@hotmail.com]
JP : Department of Astronomy, Columbia University, 550 West 20th Street, New York, NY 10027, USA [jop@astro.Columbia.edu]
MR:  Dept. of Physics, Rochester Institute of Technology,85 Lomb Memorial Drive, Rochester, NY 14623-5603, USA   [mwrsps@rit.edu]
FR:  Department of Physics, California State University, Fresno, 2345 E. San Ramon Ave., M/S MH37 Fresno, CA 93740-8031, U.S.A. [fringwal@csufresno.edu]
GR: 2007 Cedarmont Dr., Franklin, TN 37067, USA,  [georgeroberts@comcast.net]
JR: Observatory of Cantabria (CIMA, IFCA-CSIC-UC, AAC), Spain [parhelio@astrocantabria.org]
RS: 2336 Trailcrest Dr., Bozeman, MT 59718, USA [richard@theglobal.net]
WS: 6025 Calle Paraiso, Las Cruces, NM 88012, USA [starman@tbelc.org]

| Observer | Telescope | CCD | Filter |
|---|---|---|---|
| Campbell | 0.3 m SCT | SBIG ST-9XE | None |
| Foote | 0.60 m reflector | SBIG ST-8E | None |
| Garrett | 0.35 m SCT | SBIG ST10-XME | None |
| Hager | 0.25 m SCT | SBIG ST-9E | None |
| Julian | 0.3 m SCT | SBIG ST10XME | None |
| MDM telescope team | 1.3 m reflector | SITe 1024 x 1024 back illuminated detector [15], except on JD 2454611 and 2454612 when STA0500A detector was used [16] | Schott BG-38 [17] |
| Masi | 0.36 m reflector | SBIG ST8-XME | None |
| Miller | 0.35 m SCT | Starlight Xpress SXVF-H16 | None |
| Richmond | 0.3 m SCT | SBIG ST-8E | None |
| Ringwald | 0.4 m reflector | SBIG STL-11000M | None |
| Ruiz | 0.4 m SCT | SBIG ST-8XME | None |
| Sabo | 0.43 m reflector | SBIG STL-1001 | None |
| Shears | 0.1 m refractor | Starlight Xpress SXV-M7 | None |
| Stein | 0.35 m SCT | SBIG ST10XME | None |

**Table 1: Equipment used**

| Start time | Duration (h) | Observer |
|---|---|---|
| 2455022.448 | 2.01 | Shears |
| 2455022.597 | 3.65 | Richmond |
| 2455022.662 | 4.10 | Foote |
| 2455022.678 | 2.35 | Campbell |
| 2455023.576 | 1.51 | Ruiz |
| 2455023.605 | 4.97 | Garrett |
| 2455023.680 | 3.72 | Stein |
| 2455023.705 | 4.94 | Sabo |
| 2455024.595 | 1.30 | Ruiz |
| 2455024.672 | 4.13 | Stein |
| 2455025.337 | 3.50 | Masi |
| 2455025.583 | 4.03 | Richmond |
| 2455025.631 | 5.35 | Julian |
| 2455025.657 | 4.22 | Stein |
| 2455025.738 | 3.86 | Ringwald |
| 2455026.314 | 3.94 | Masi |
| 2455026.574 | 4.27 | Richmond |
| 2455026.577 | 1.20 | Hager |
| 2455026.650 | 4.44 | Stein |
| 2455026.704 | 4.58 | Ringwald |
| 2455027.572 | 4.32 | Richmond |
| 2455027.678 | 5.16 | Ringwald |
| 2455027.717 | 3.14 | Sabo |
| 2455028.432 | 3.48 | Miller |
| 2455028.704 | 0.89 | Sabo |
| 2455029.715 | 4.13 | Ringwald |
| 2455030.493 | 2.02 | Miller |
| 2455030.689 | 4.68 | Ringwald |
| 2455031.725 | 4.08 | Sabo |
| 2455032.567 | 2.40 | Richmond |
| 2455032.683 | 4.70 | Ringwald |
| 2455033.685 | 4.54 | Ringwald |
| 2455034.690 | 4.32 | Ringwald |
| 2455038.530 | 1.13 | Ruiz |
| 2455081.358 | 1.06 | Ruiz |

(a) Photometry during the July 2009 outburst

| Start time | Duration (h) |
|---|---|
| 2453849.898 | 2.62 |
| 2453850.649 | 8.54 |
| 2453851.648 | 8.64 |
| 2453852.639 | 5.88 |
| 2453876.651 | 1.22 |
| 2454179.820 | 4.97 |
| 2454611.811 | 3.24 |
| 2454612.651 | 7.80 |

(b) Photometry during quiescence (1.3 m MDM telescope)

**Table 2 : Log of time-series photometry**

| Orbital cycle number | Eclipse minimum (HJD) | O-C (orbital cycles) | Error (orbital cycles) | Eclipse depth (mag) | Eclipse duration FWHM (min) |
|---|---|---|---|---|---|
| 0 | 2453849.94908 | -0.0001 | 0.0059 | | |
| 1 | 2453850.00684 | -0.0196 | 0.0016 | | |
| 12 | 2453850.65615 | 0.0026 | 0.0020 | | |
| 13 | 2453850.71498 | 0.0013 | 0.0018 | | |
| 14 | 2453850.77384 | 0.0004 | 0.0035 | | |
| 15 | 2453850.83284 | 0.0020 | 0.0021 | | |
| 16 | 2453850.89165 | 0.0002 | 0.0023 | | |
| 17 | 2453850.95066 | 0.0020 | 0.0022 | | |
| 29 | 2453851.65760 | 0.0024 | 0.0024 | | |
| 30 | 2453851.71630 | -0.0012 | 0.0023 | | |
| 32 | 2453851.83417 | -0.0002 | 0.0028 | | |
| 33 | 2453851.89333 | 0.0040 | 0.0023 | | |
| 34 | 2453851.95249 | 0.0083 | 0.0023 | | |
| 46 | 2453852.65899 | 0.0013 | 0.0031 | | |
| 47 | 2453852.71800 | 0.0029 | 0.0026 | | |
| 48 | 2453852.77661 | -0.0021 | 0.0018 | | |
| 49 | 2453852.83568 | 0.0006 | 0.0023 | | |
| 454 | 2453876.69411 | 0.0023 | 0.0007 | | |
| 5600 | 2454179.84201 | -0.0008 | 0.0016 | | |
| 5601 | 2454179.90009 | -0.0149 | 0.0016 | | |
| 5603 | 2454180.01900 | 0.0036 | 0.0019 | | |
| 12933 | 2454611.82513 | 0.0000 | 0.0020 | | |
| 12934 | 2454611.88410 | 0.0010 | 0.0020 | | |
| 12935 | 2454611.94324 | 0.0050 | 0.0029 | | |
| 12948 | 2454612.70875 | -0.0003 | 0.0032 | | |
| 12949 | 2454612.76794 | 0.0043 | 0.0020 | | |
| 12950 | 2454612.82684 | 0.0043 | 0.0030 | | |
| 12951 | 2454612.88550 | 0.0000 | 0.0021 | | |
| 12952 | 2454612.94455 | 0.0023 | 0.0045 | | |
| 19904 | 2455022.48416 | 0.0202 | 0.0008 | 0.93 | 10.50 |
| 19906 | 2455022.60181 | 0.0173 | 0.0043 | 0.93 | 10.40 |
| 19907 | 2455022.66054 | 0.0143 | 0.0040 | 0.94 | 10.35 |
| 19908 | 2455022.71933 | 0.0123 | 0.0026 | 0.95 | 10.35 |
| 19908 | 2455022.71925 | 0.0109 | 0.0063 | 0.96 | 9.90 |
| 19908 | 2455022.71904 | 0.0073 | 0.0037 | 0.94 | 10.35 |
| 19909 | 2455022.77860 | 0.0183 | 0.0053 | 0.92 | 9.60 |
| 19923 | 2455023.60296 | 0.0121 | 0.0034 | 0.89 | 8.25 |
| 19924 | 2455023.66137 | 0.0035 | 0.0047 | 0.93 | 9.75 |
| 19925 | 2455023.72090 | 0.0141 | 0.0016 | 0.94 | 10.50 |
| 19925 | 2455023.72081 | 0.0125 | 0.0068 | 0.93 | 9.60 |
| 19926 | 2455023.77969 | 0.0121 | 0.0037 | 1.00 | 9.75 |
| 19926 | 2455023.77951 | 0.0089 | 0.0032 | 0.99 | 10.05 |
| 19926 | 2455023.77946 | 0.0081 | 0.0035 | 0.98 | 9.60 |
| 19927 | 2455023.83812 | 0.0039 | 0.0024 | 0.82 | 9.30 |
| 19928 | 2455023.89723 | 0.0073 | 0.0090 | 0.85 | 9.75 |
| 19940 | 2455024.60310 | -0.0104 | 0.0085 | 0.87 | 10.35 |
| 19942 | 2455024.72186 | 0.0056 | 0.0057 | 1.08 | 10.35 |
| 19943 | 2455024.78086 | 0.0072 | 0.0038 | 1.10 | 9.60 |
| 19944 | 2455024.83943 | 0.0014 | 0.0052 | 1.13 | 9.75 |
| 19953 | 2455025.37019 | 0.0111 | 0.0043 | 0.98 | 8.85 |
| 19954 | 2455025.42911 | 0.0113 | 0.0028 | 0.96 | 8.40 |
| 19957 | 2455025.60577 | 0.0101 | 0.0035 | 0.96 | 8.70 |
| 19958 | 2455025.66470 | 0.0105 | 0.0134 | 0.97 | 8.25 |
| 19958 | 2455025.66476 | 0.0114 | 0.0041 | 0.96 | 8.70 |
| 19958 | 2455025.66454 | 0.0077 | 0.0046 | 0.96 | 10.05 |
| 19959 | 2455025.72346 | 0.0079 | 0.0013 | 0.99 | 10.20 |
| 19959 | 2455025.72353 | 0.0091 | 0.0020 | 0.98 | 9.60 |
| 19959 | 2455025.72335 | 0.0061 | 0.0033 | 1.00 | 9.90 |
| 19960 | 2455025.78253 | 0.0107 | 0.0013 | 0.96 | 9.60 |
| 19960 | 2455025.78255 | 0.0109 | 0.0013 | 0.95 | 6.15 |
| 19960 | 2455025.78271 | 0.0137 | 0.0018 | 0.97 | 9.00 |
| 19961 | 2455025.84149 | 0.0115 | 0.0022 | 1.00 | 8.55 |
| 19961 | 2455025.84167 | 0.0146 | 0.0041 | 1.01 | 7.50 |
| 19970 | 2455026.37155 | 0.0093 | 0.0045 | 0.63 | 9.75 |
| 19971 | 2455026.43029 | 0.0064 | 0.0048 | 0.60 | 6.30 |
| 19974 | 2455026.60730 | 0.0112 | 0.0019 | ND | ND |
| 19975 | 2455026.66539 | -0.0026 | 0.0058 | 1.07 | 8.70 |
| 19976 | 2455026.72392 | -0.0092 | 0.0196 | 1.24 | 10.95 |
| 19976 | 2455026.72392 | -0.0091 | 0.0055 | 1.23 | 10.05 |
| 19976 | 2455026.72478 | 0.0054 | 0.0111 | 1.25 | 9.75 |
| 19977 | 2455026.78257 | -0.0135 | 0.0095 | 1.29 | 10.95 |
| 19991 | 2455027.60878 | 0.0116 | 0.0067 | 1.04 | 8.55 |
| 19992 | 2455027.66745 | 0.0075 | 0.0053 | 1.07 | 7.35 |
| 19993 | 2455027.72653 | 0.0105 | 0.0063 | 1.11 | 8.55 |
| 19993 | 2455027.72639 | 0.0081 | 0.0044 | 1.10 | 8.10 |
| 19993 | 2455027.72649 | 0.0097 | 0.0025 | 1.12 | 9.45 |
| 19994 | 2455027.78564 | 0.0138 | 0.0038 | 1.10 | 8.85 |
| 19995 | 2455027.84456 | 0.0141 | 0.0050 | 1.13 | 7.50 |
| 19995 | 2455027.84448 | 0.0126 | 0.0484 | 1.12 | 7.35 |
| 20006 | 2455028.49234 | 0.0102 | 0.0362 | 1.12 | 7.50 |
| 20007 | 2455028.55092 | 0.0046 | 0.0036 | 1.07 | 7.80 |
| 20010 | 2455028.72746 | 0.0013 | 0.0214 | ND | ND |
| 20027 | 2455029.72931 | 0.0081 | 0.0021 | 1.56 | 7.35 |
| 20028 | 2455029.78814 | 0.0067 | 0.0030 | 1.61 | 7.80 |
| 20029 | 2455029.84733 | 0.0115 | 0.0038 | 1.51 | 6.00 |
| 20040 | 2455030.49545 | 0.0134 | 0.0179 | 1.56 | 7.05 |
| 20041 | 2455030.55408 | 0.0087 | 0.0040 | 1.70 | 6.60 |
| 20044 | 2455030.73050 | 0.0034 | 0.0023 | 1.52 | 5.70 |
| 20045 | 2455030.78955 | 0.0057 | 0.0014 | 1.47 | 6.00 |
| 20046 | 2455030.84850 | 0.0066 | 0.0027 | 1.50 | 5.40 |
| 20061 | 2455031.73191 | 0.0026 | 0.0107 | 1.47 | 5.85 |
| 20062 | 2455031.79041 | -0.0043 | 0.0073 | 1.58 | 5.70 |
| 20063 | 2455031.84955 | -0.0005 | 0.0184 | 1.22 | 4.35 |
| 20078 | 2455032.73313 | -0.0015 | 0.0082 | 1.69 | 4.05 |
| 20078 | 2455032.73321 | -0.0001 | 0.0163 | 1.87 | 4.20 |
| 20078 | 2455032.73313 | -0.0015 | 0.0200 | 1.98 | 3.90 |
| 20079 | 2455032.79204 | -0.0015 | 0.0171 | 1.95 | 3.60 |
| 20080 | 2455032.85150 | 0.0079 | 0.0079 | 2.05 | 3.30 |
| 20080 | 2455032.85207 | 0.0176 | 0.0210 | 1.85 | 3.15 |
| 20095 | 2455033.73521 | 0.0090 | 0.0163 | 2.02 | 3.00 |
| 20096 | 2455033.79404 | 0.0076 | 0.0171 | 2.06 | 2.90 |
| 20097 | 2455033.85207 | -0.0073 | 0.0210 | 2.18 | 3.10 |
| 20112 | 2455034.73613 | -0.0002 | 0.0200 | ND | ND |
| 20904 | 2455081.39294 | 0.0086 | 0.0286 | ND | ND |

**Table 3: Eclipse minimum times, depth and duration**
Note: depth and duration was only measured during the outburst.
ND = not determined

| Superhump cycle | Superhump maximum (HJD) | O-C (superhump cycles) | Error (superhump cycles) |
|---|---|---|---|
| 0 | 2455022.5092 | 0.0160 | 0.0124 |
| 2 | 2455022.6288 | 0.0009 | 0.0087 |
| 3 | 2455022.6890 | -0.0004 | 0.0082 |
| 3 | 2455022.6888 | -0.0034 | 0.0106 |
| 3 | 2455022.6892 | 0.0026 | 0.0027 |
| 4 | 2455022.7489 | -0.0066 | 0.0079 |
| 4 | 2455022.7492 | -0.0023 | 0.0079 |
| 5 | 2455022.8090 | -0.0097 | 0.0074 |
| 37 | 2455024.7391 | 0.0078 | 0.0088 |
| 38 | 2455024.8003 | 0.0231 | 0.0122 |
| 47 | 2455025.3404 | -0.0153 | 0.0086 |
| 48 | 2455025.4001 | -0.0246 | 0.0154 |
| 49 | 2455025.4579 | -0.0641 | 0.0250 |
| 52 | 2455025.6396 | -0.0512 | 0.0144 |
| 52 | 2455025.6394 | -0.0535 | 0.0212 |
| 53 | 2455025.7027 | -0.0055 | 0.0187 |
| 53 | 2455025.7063 | 0.0523 | 0.0229 |
| 54 | 2455025.7645 | 0.0189 | 0.0119 |
| 54 | 2455025.7653 | 0.0315 | 0.0179 |
| 54 | 2455025.7653 | 0.0146 | 0.0108 |
| 55 | 2455025.8260 | 0.0222 | 0.0270 |
| 55 | 2455025.8256 | 0.0159 | 0.0382 |
| 55 | 2455025.8287 | 0.0643 | 0.0063 |
| 84 | 2455027.5806 | 0.1253 | 0.0092 |
| 85 | 2455027.6423 | 0.1477 | 0.0226 |
| 86 | 2455027.7038 | 0.1677 | 0.0075 |
| 86 | 2455027.7042 | 0.1739 | 0.0114 |
| 87 | 2455027.7653 | 0.1859 | 0.0069 |
| 88 | 2455027.8247 | 0.1730 | 0.0094 |
| 88 | 2455027.8244 | 0.1675 | 0.0080 |
| 89 | 2455027.8862 | 0.1913 | 0.0090 |
| 120 | 2455029.7483 | 0.0870 | 0.0074 |
| 122 | 2455029.8695 | 0.0982 | 0.0075 |
| 136 | 2455030.7095 | 0.0357 | 0.0348 |
| 137 | 2455030.7717 | 0.0656 | 0.0343 |
| 138 | 2455030.8370 | 0.1457 | 0.0367 |
| 169 | 2455032.7243 | 0.4425 | 0.0081 |

**Table 4: Superhump maximum times**

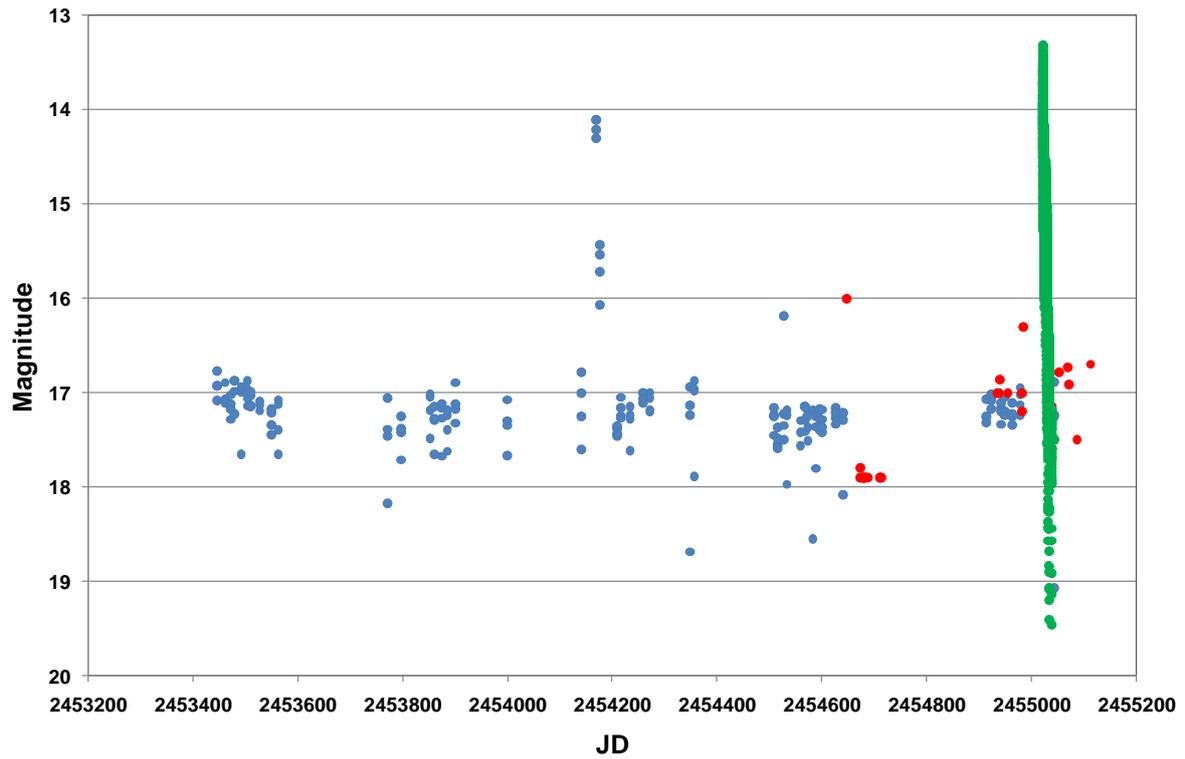

**Figure 1: Light curve of SDSS 1502 between 2005 Mar 15 and 2009 Oct 10**
Data sources: blue = CRTS, red = the authors' (discrete measurements), green = the authors (time-series photometry from the July 2009 outburst as presented in Figure 2)

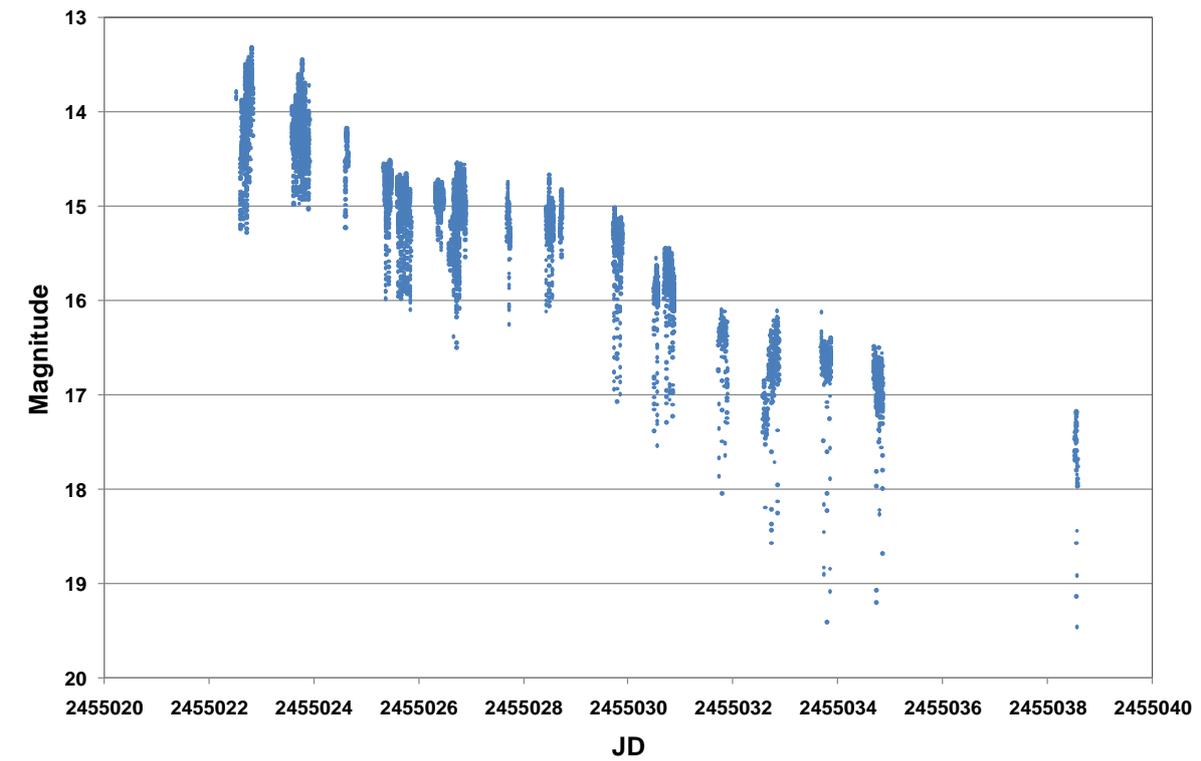

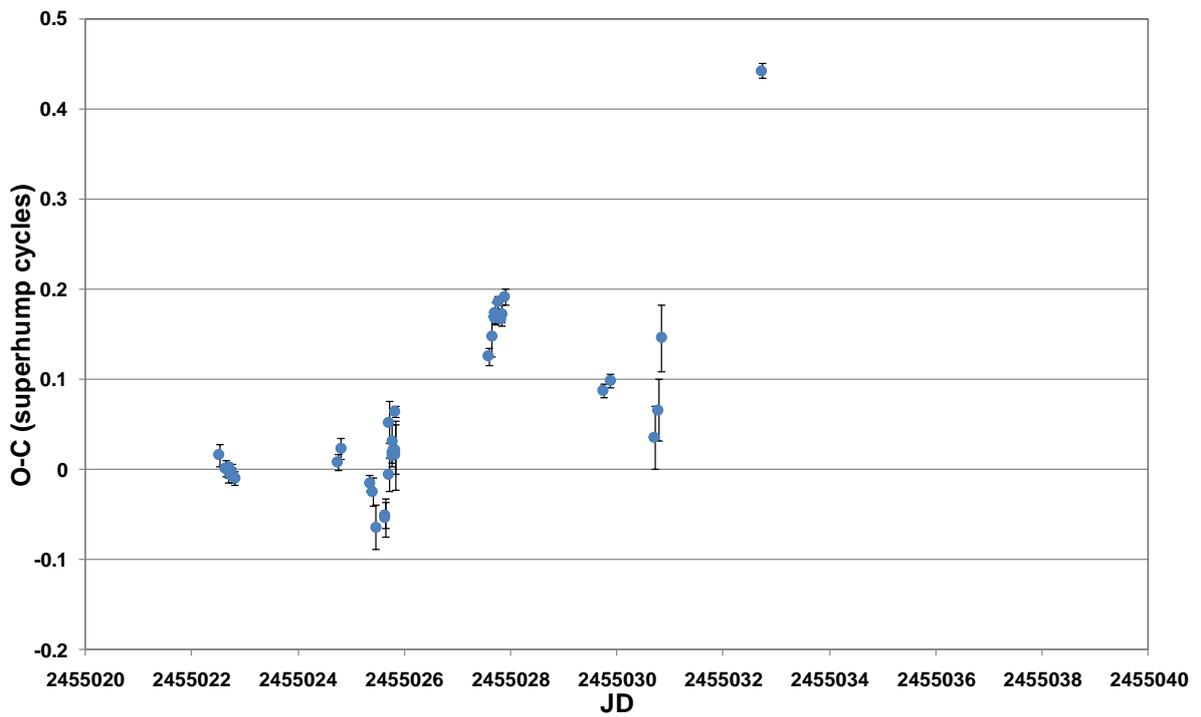

**Figure 2: Light curve of the outburst (top) and O-C diagram of superhump maxima relative to the ephemeris in equation 2 (bottom)**

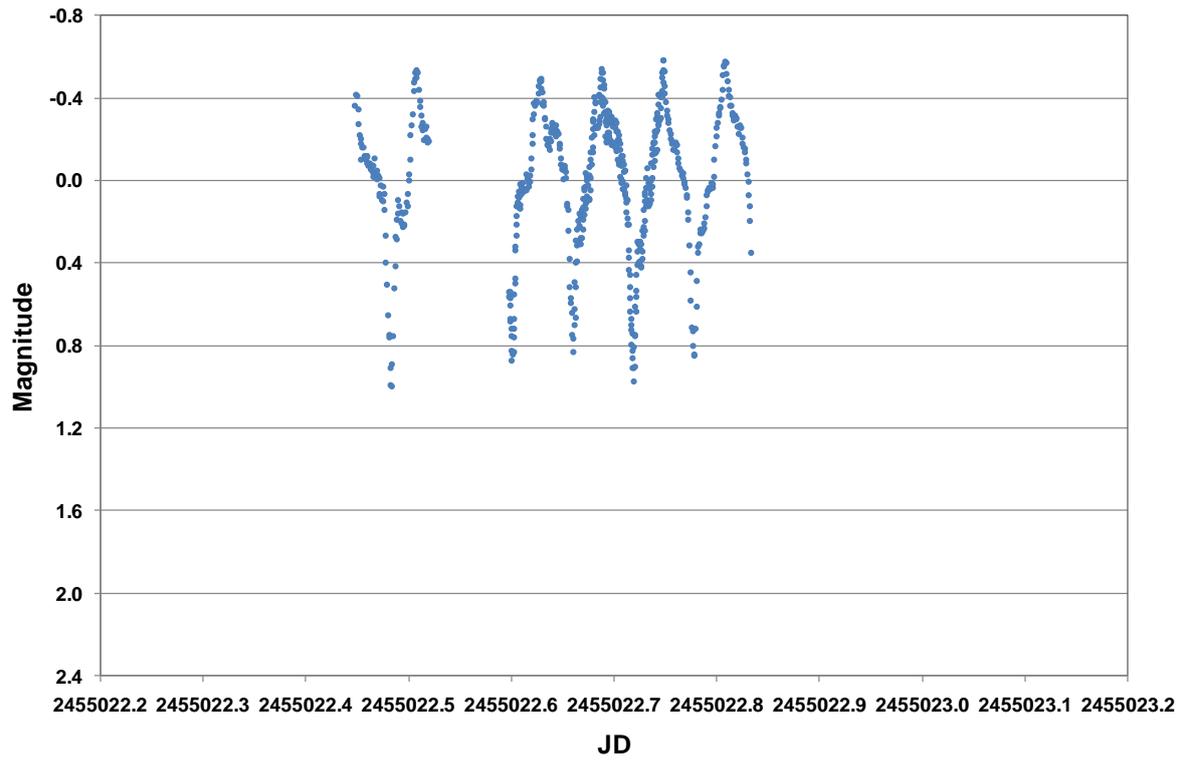
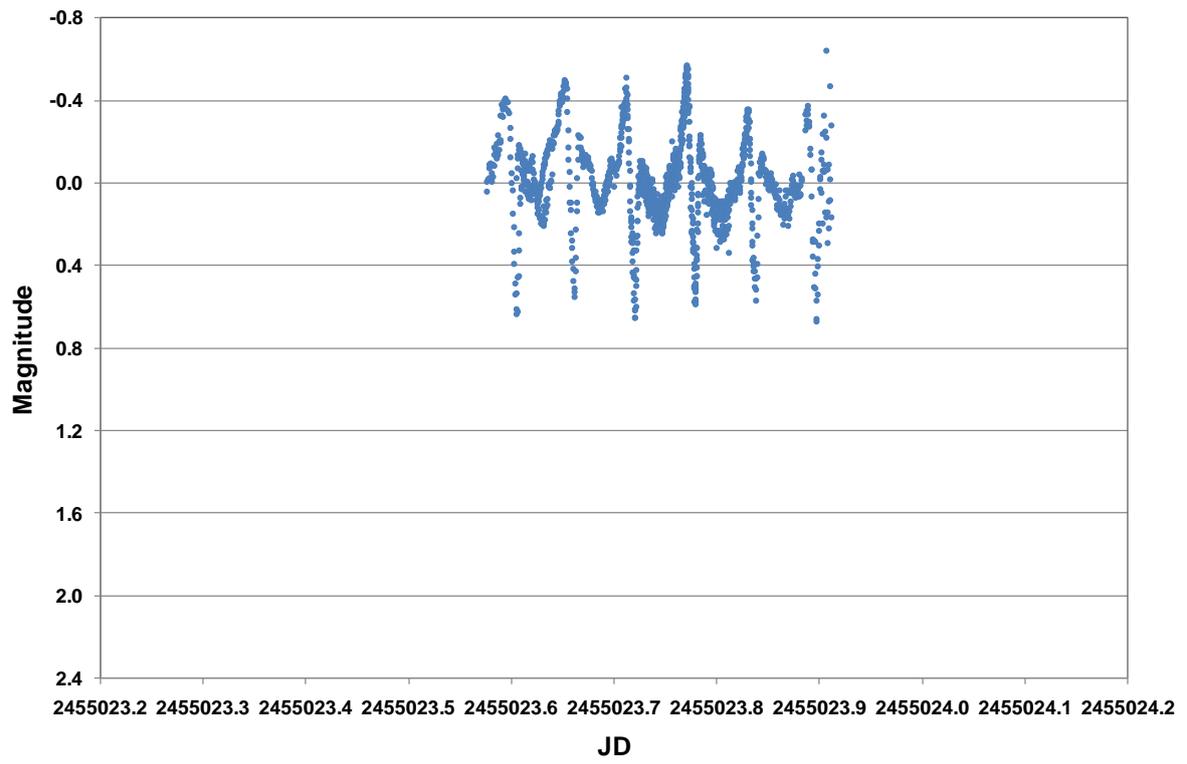

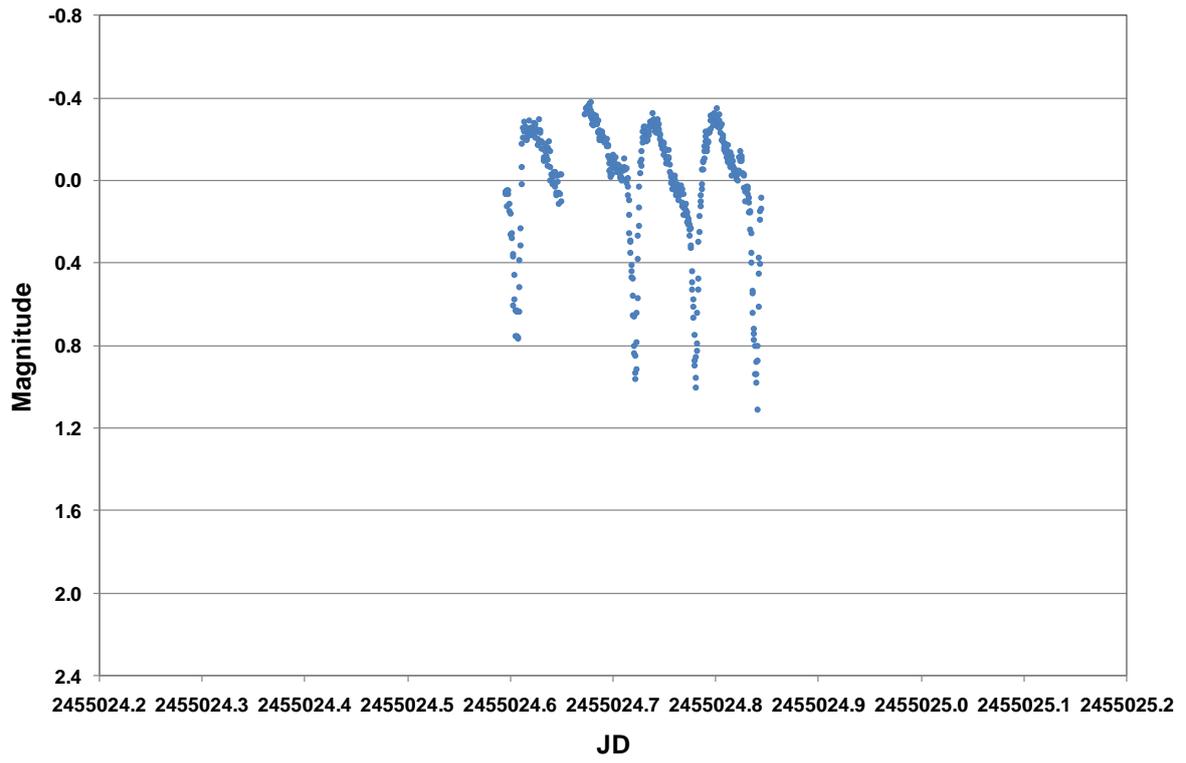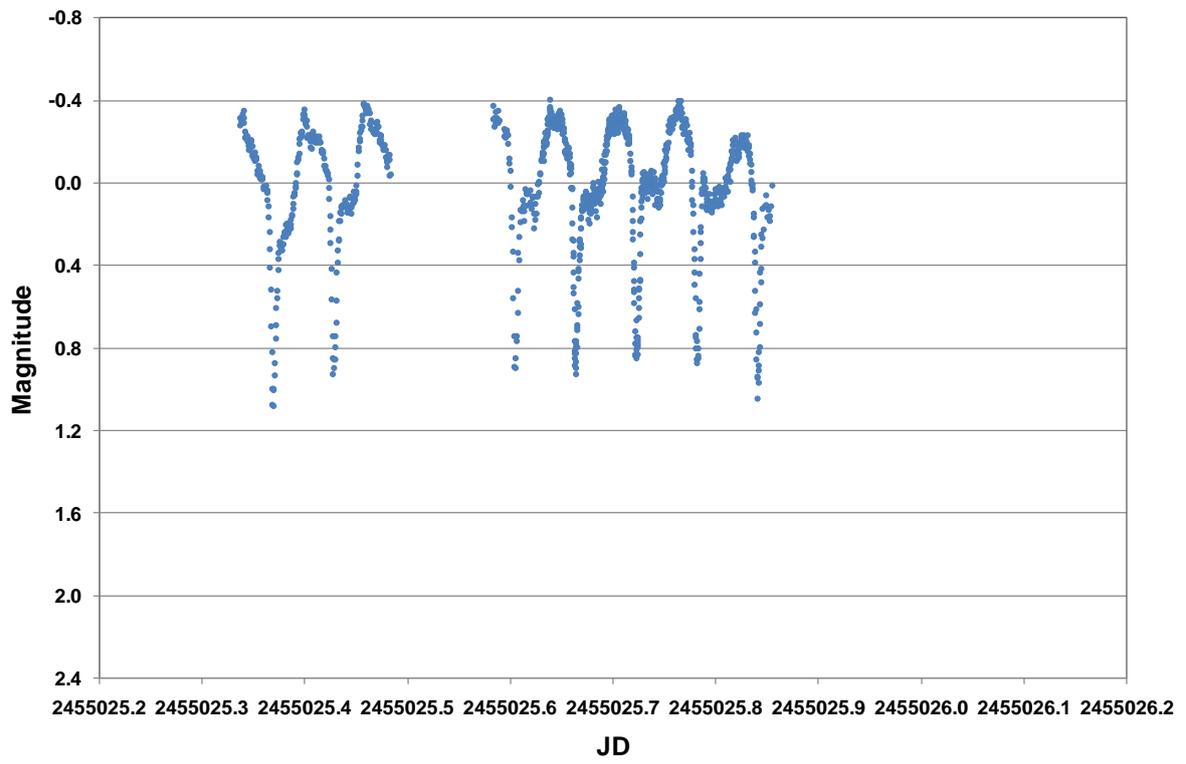

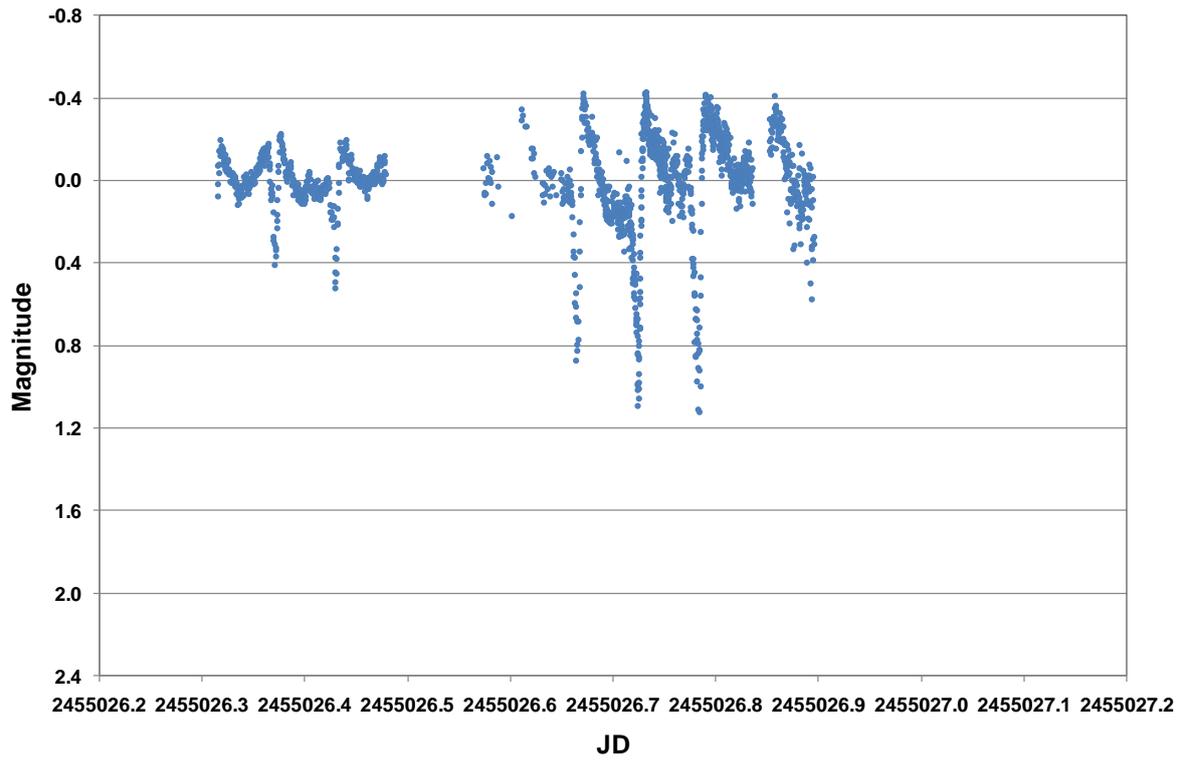
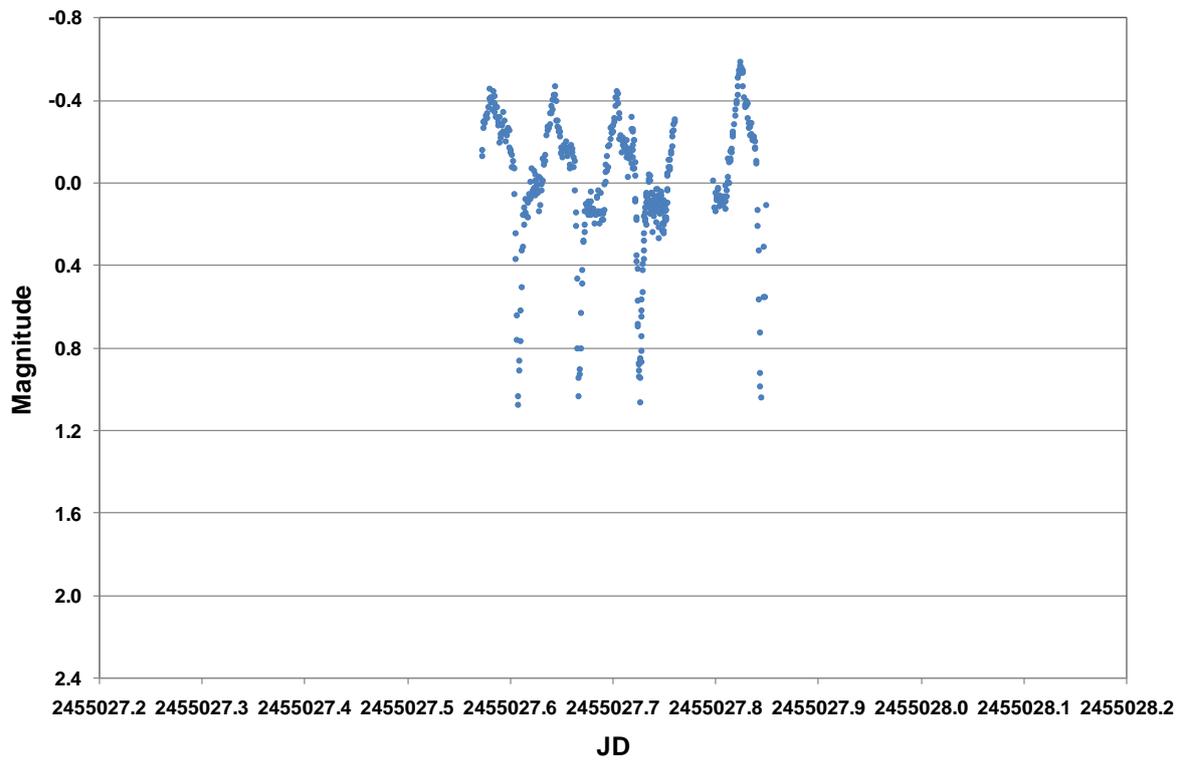

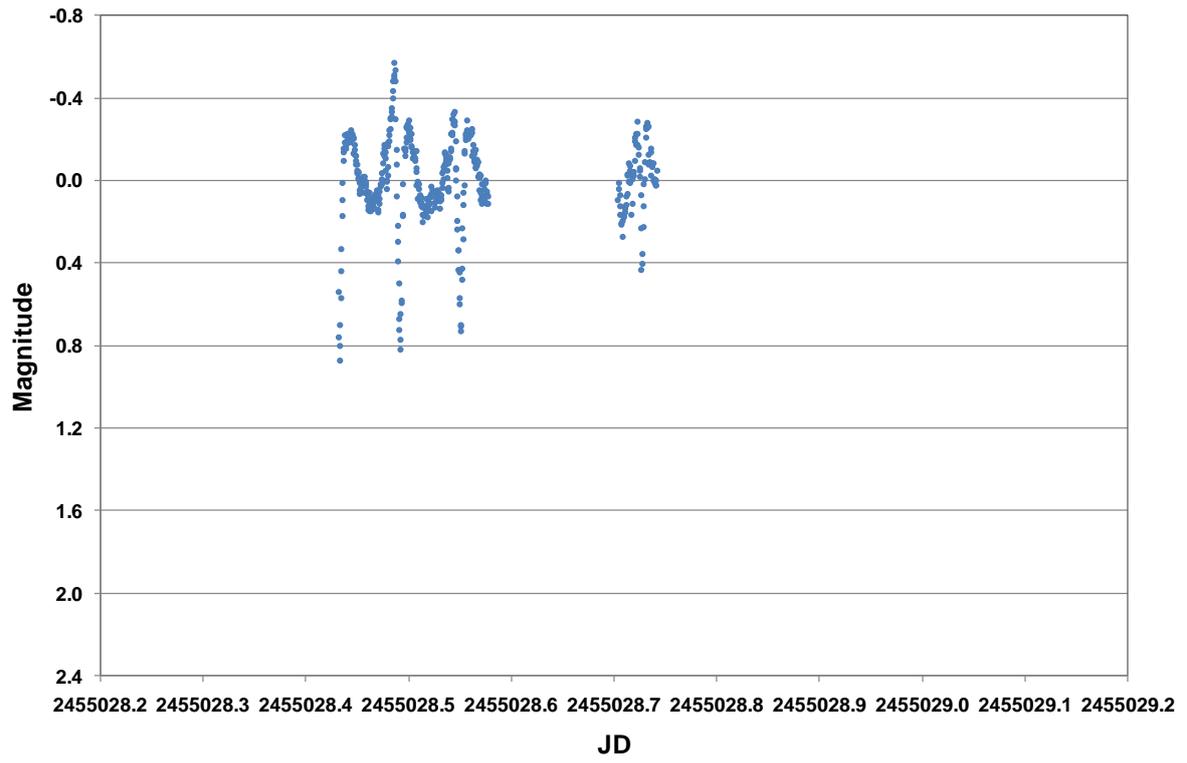

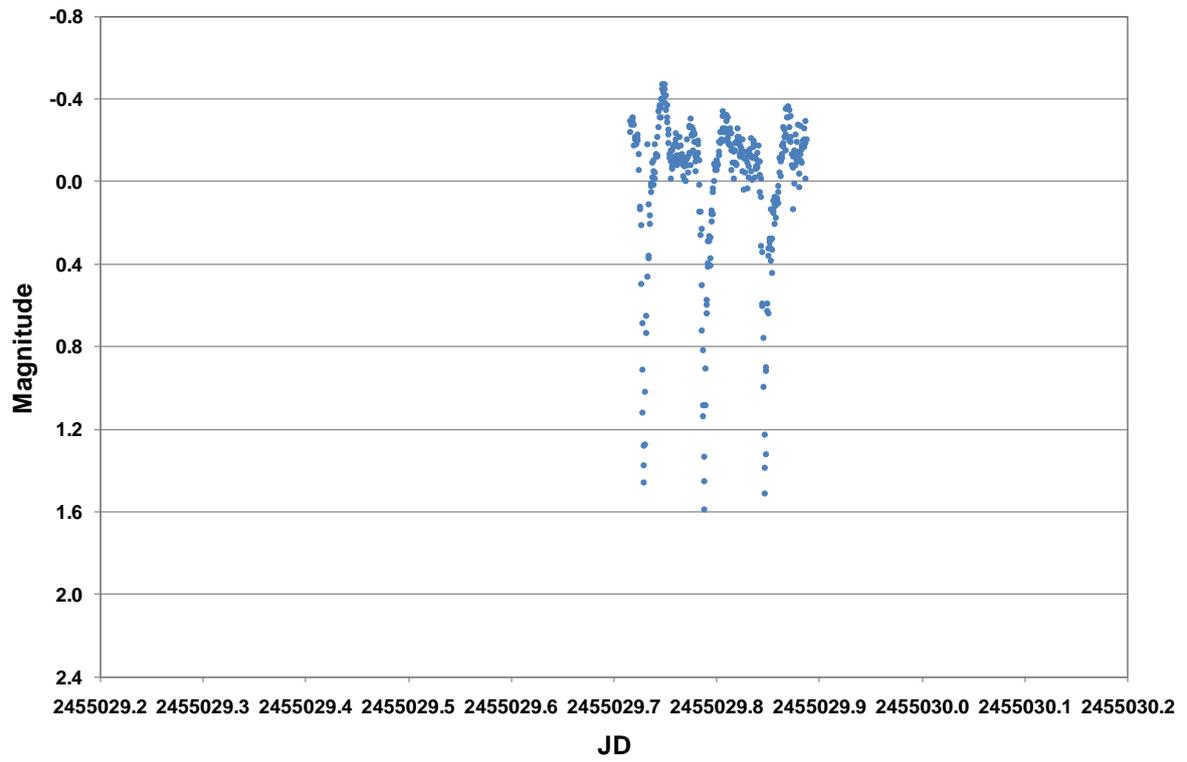

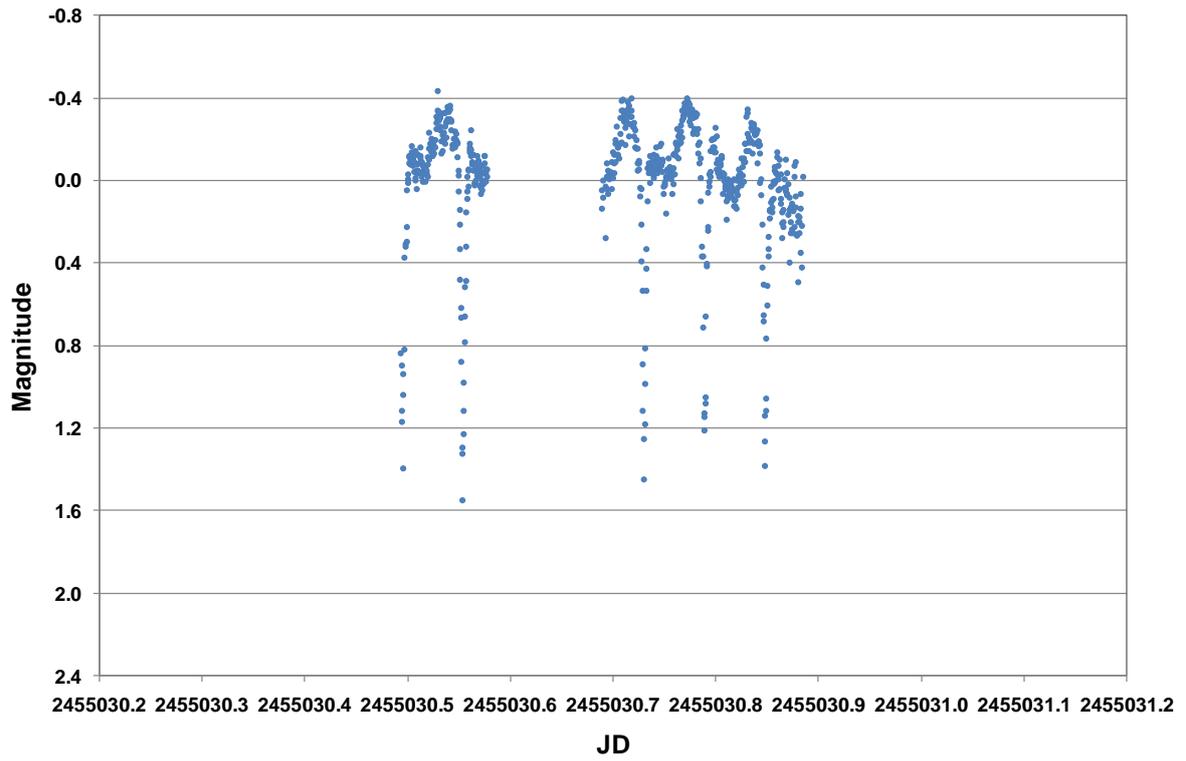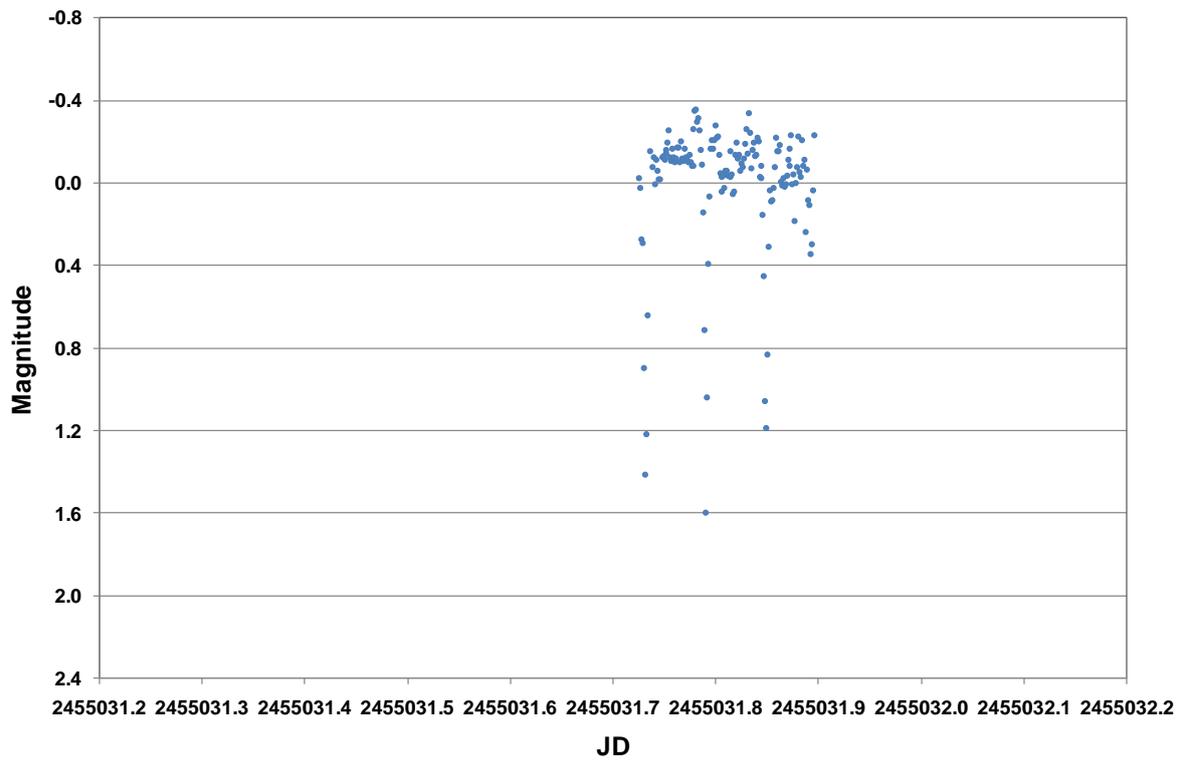

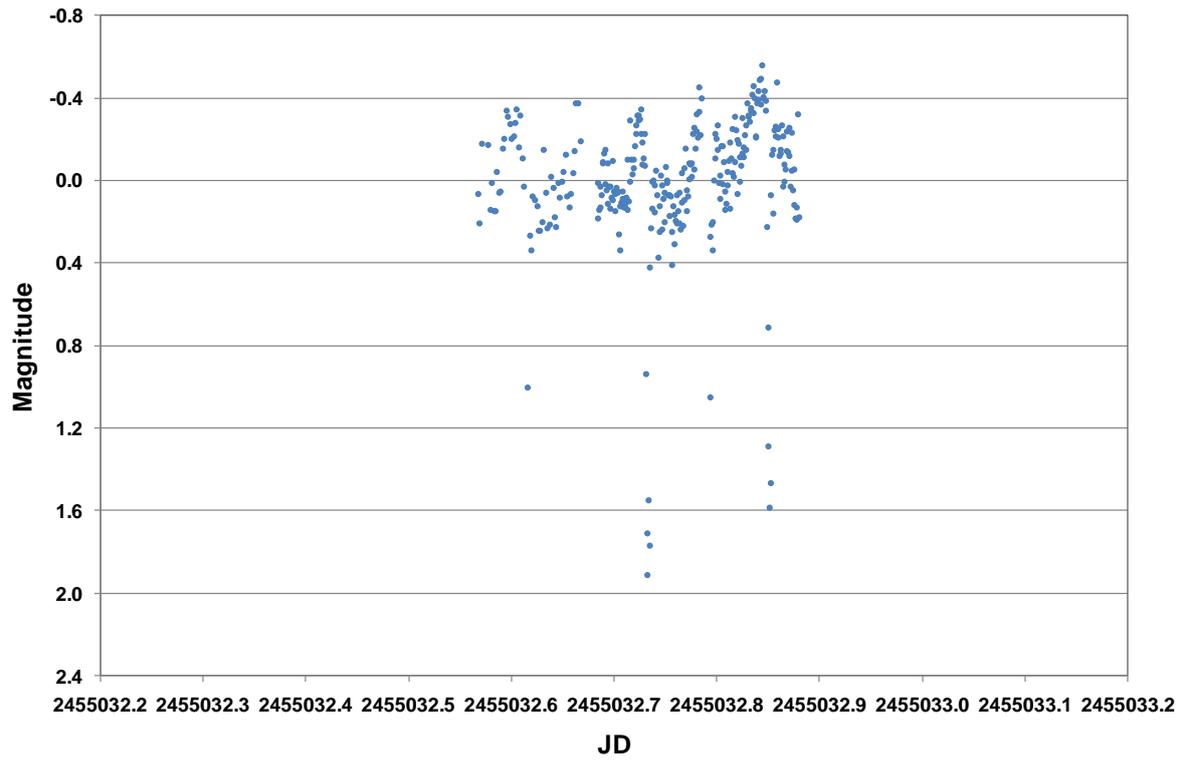

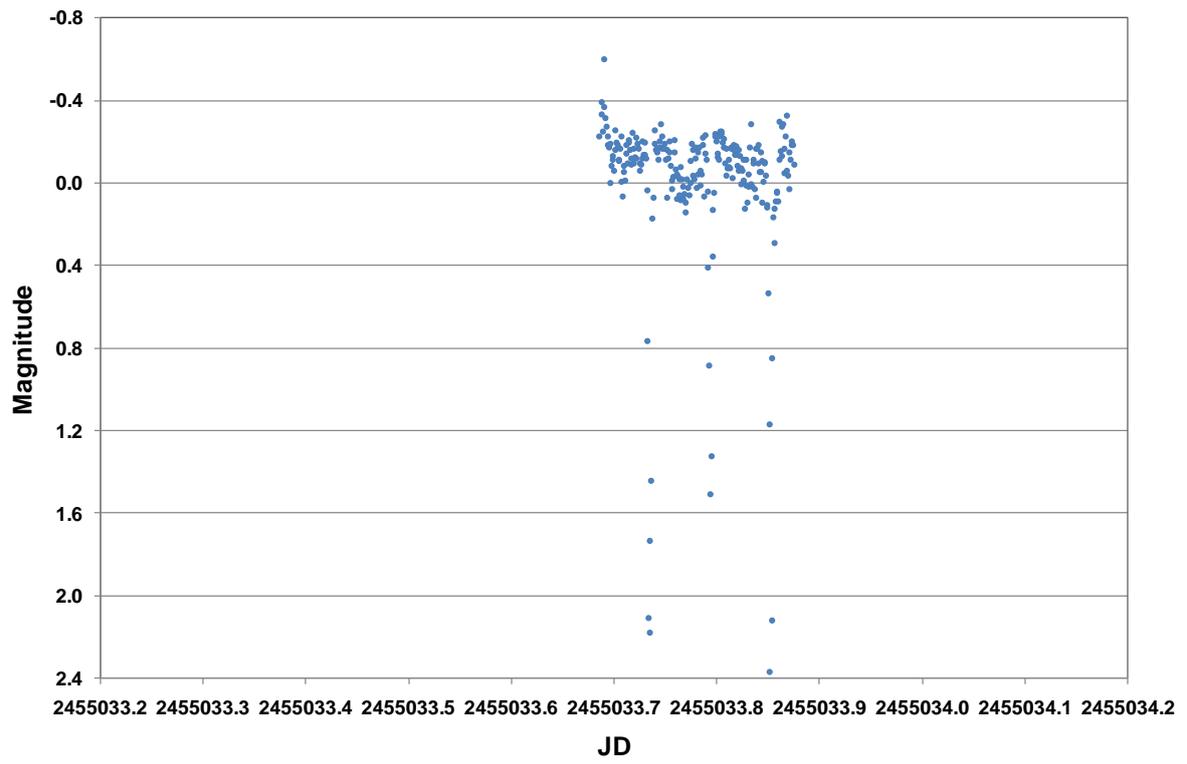

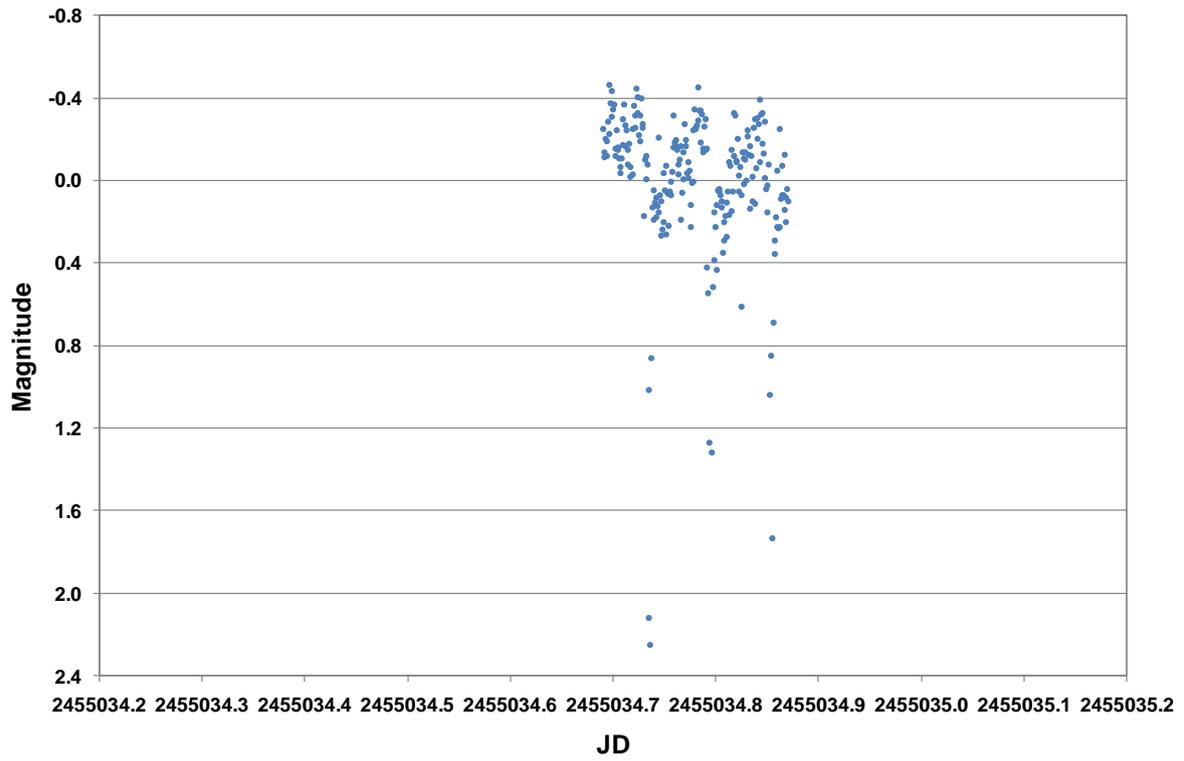
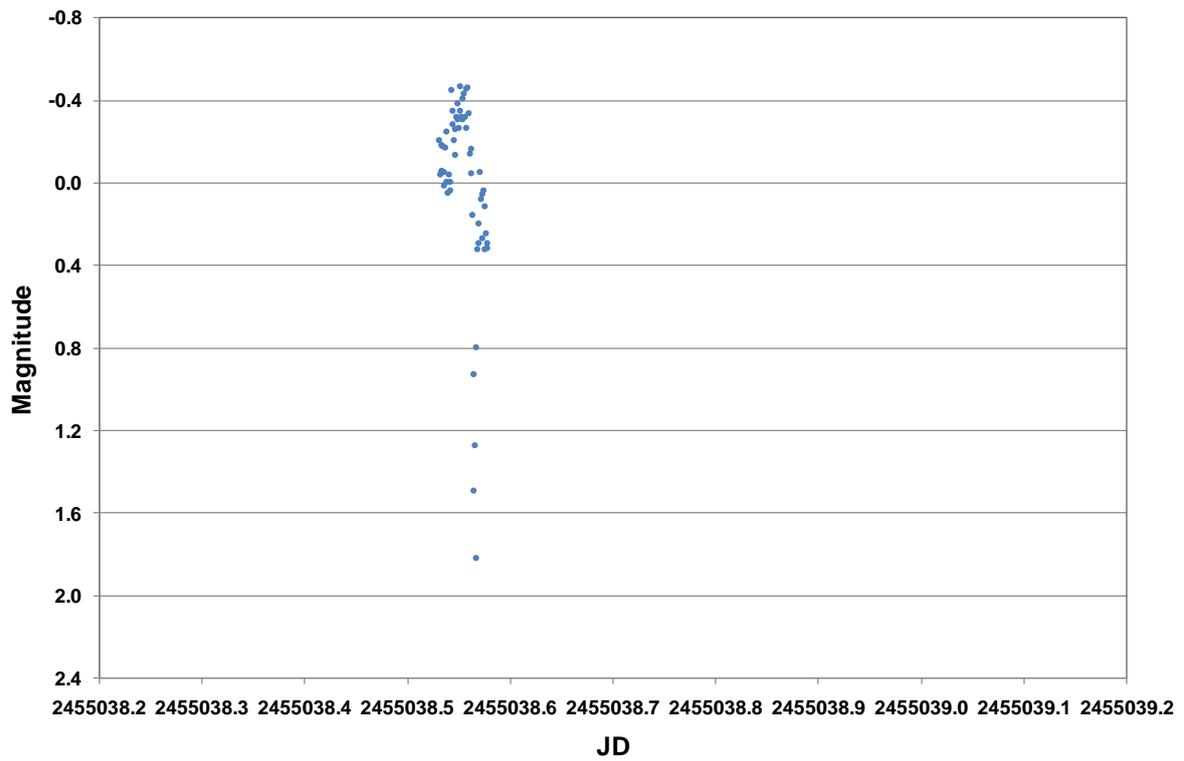

**Figure 3: Time series photometry during the outburst of SDSS 1502**
*Note: Figure 3 extends over this page and the previous 6 pages to make the plots more easily legible to the referees*

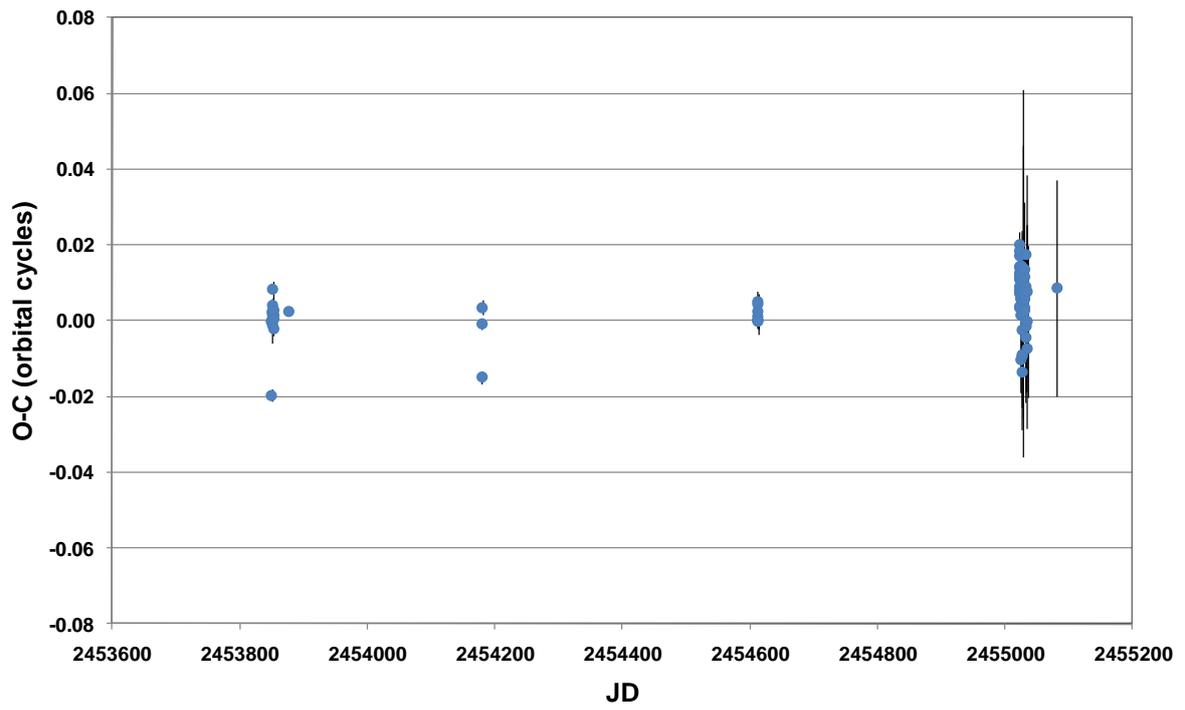

**Figure 4: O-C residuals for the eclipses**

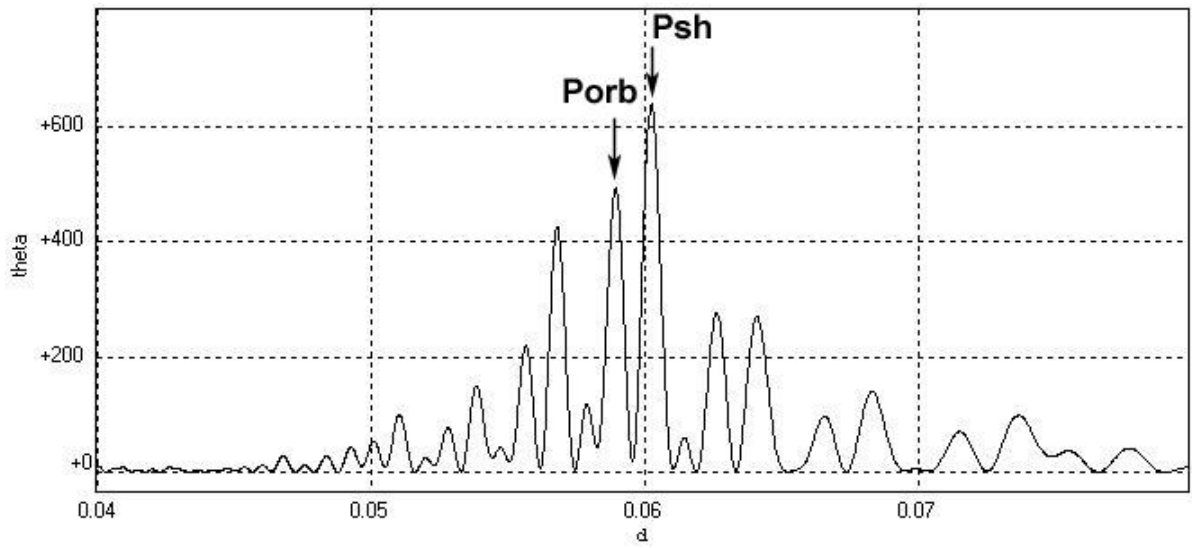

**Figure 5: DCDFT power spectrum of data from JD 2455022 to 2455025**

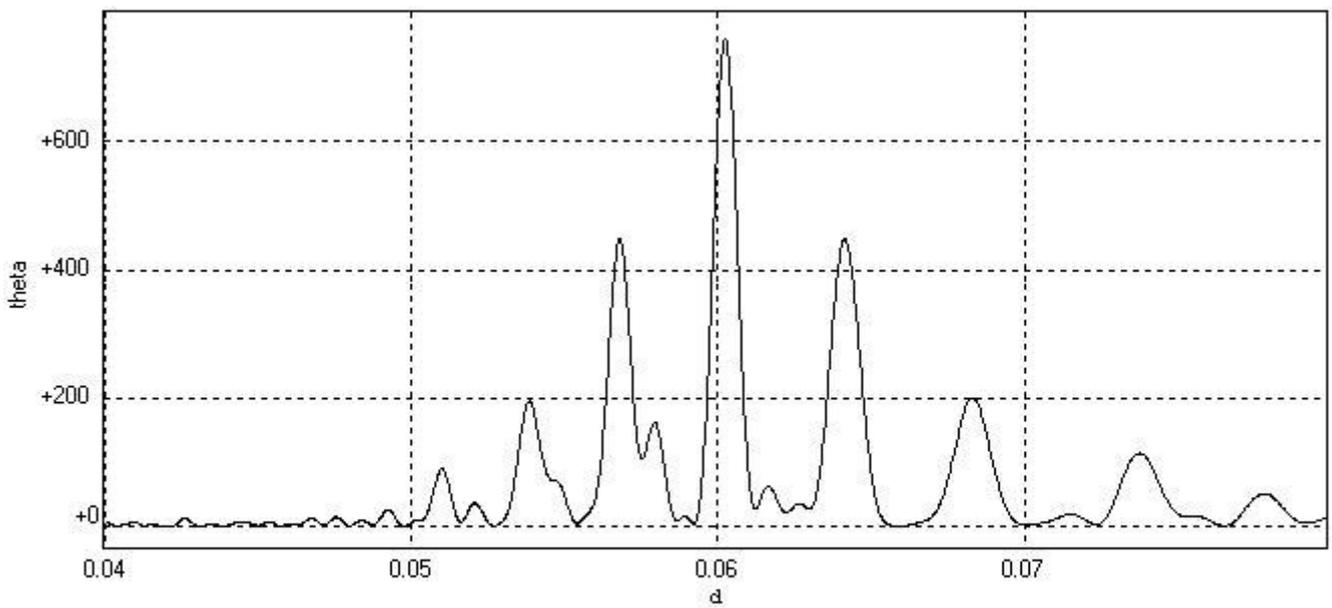

**Figure 6: DCDFT power spectrum of after pre-whitening with $P_{orb}$**

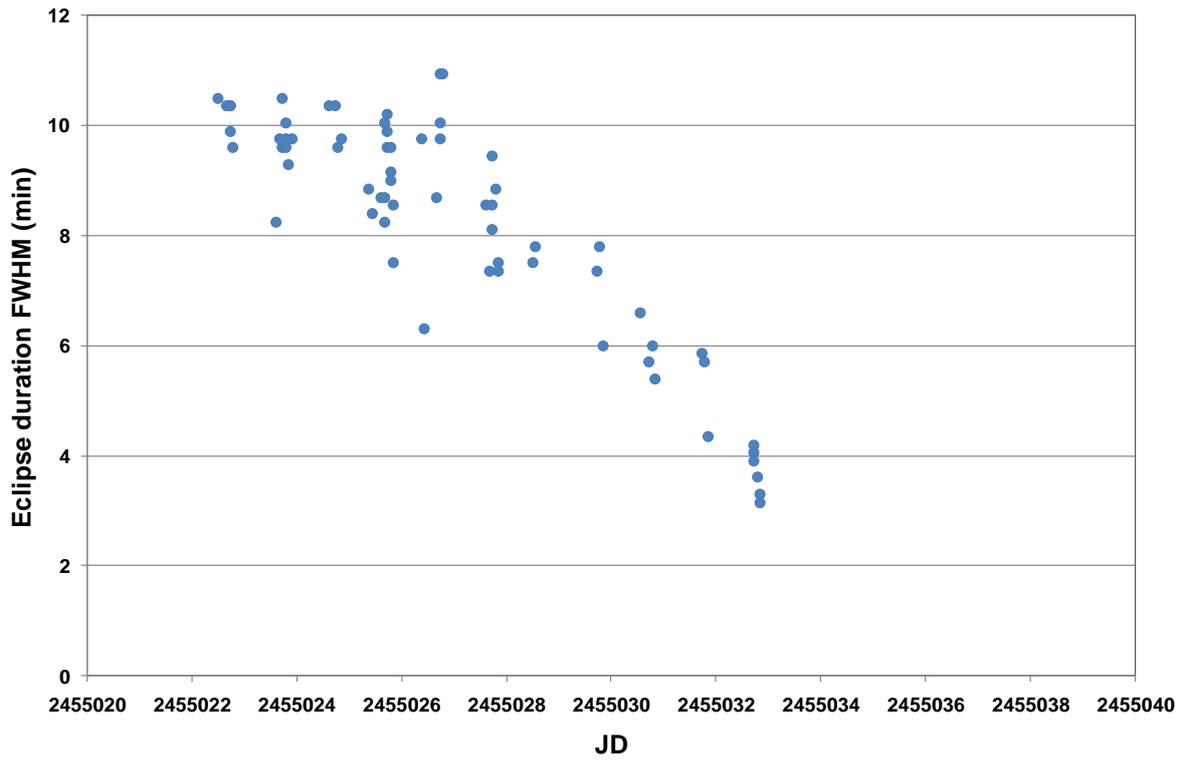

**Figure 7: Duration of eclipses during the outburst**

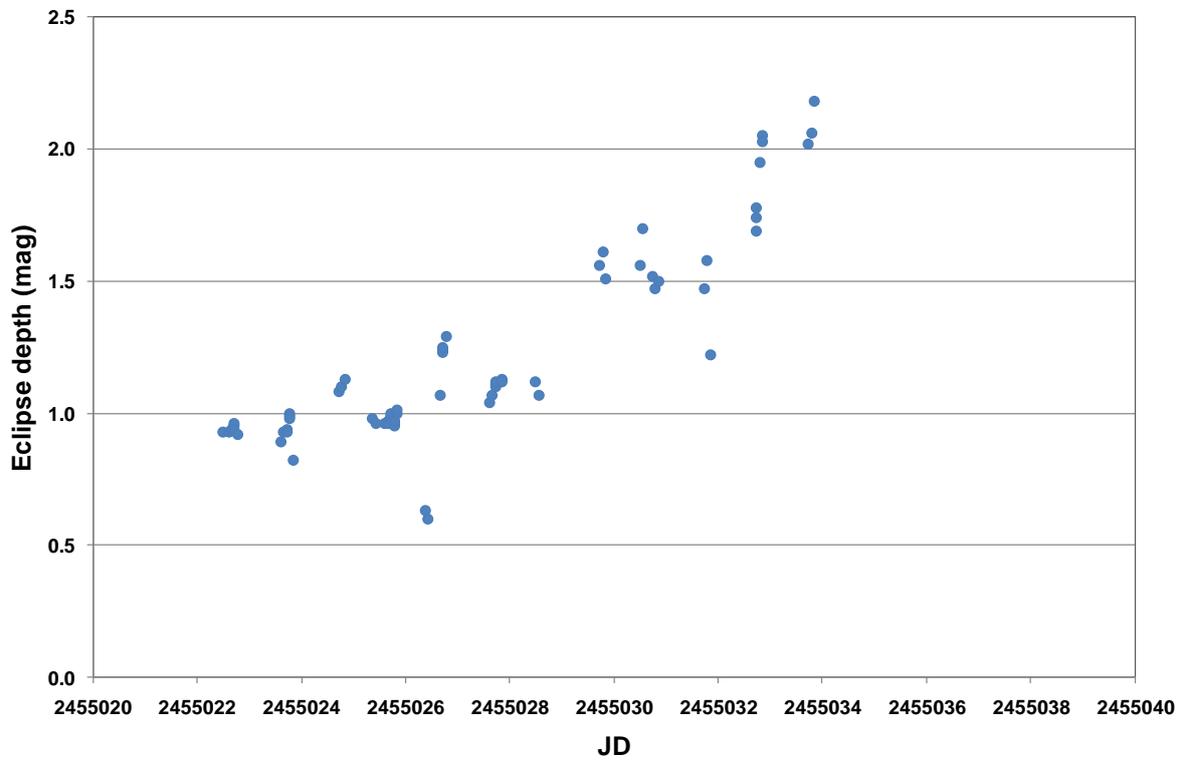

**Figure 8: Depth of eclipses during the outburst**

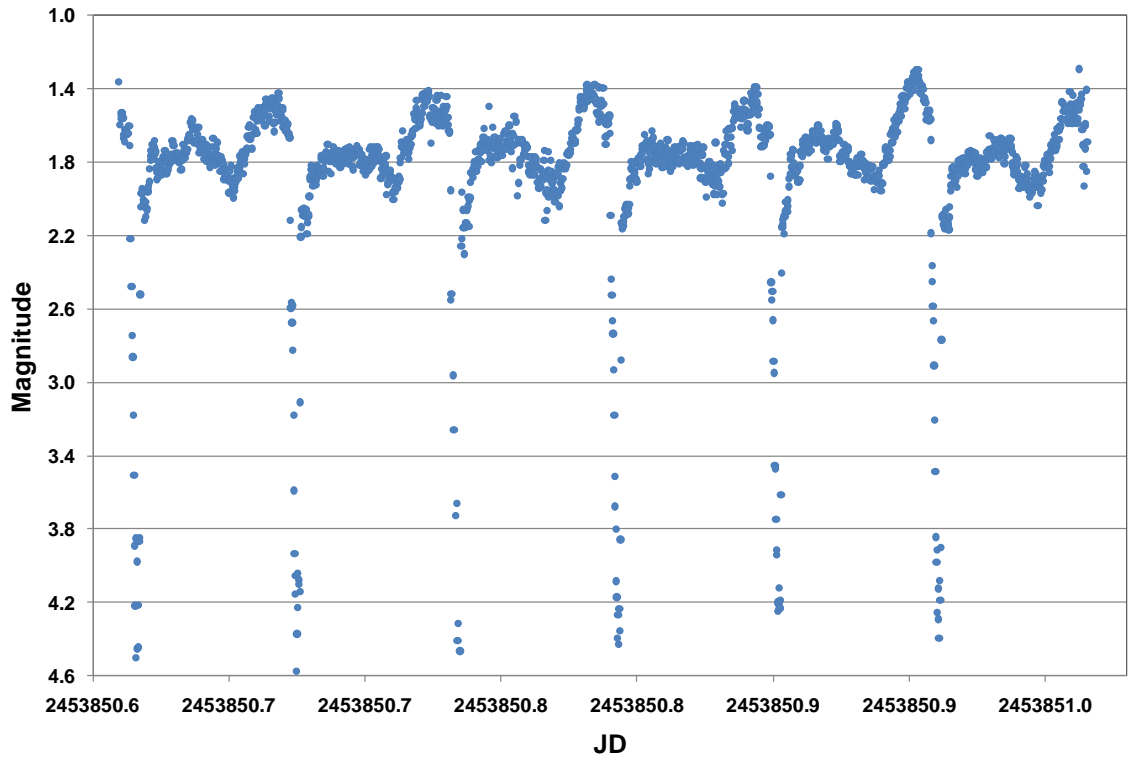

**Figure 9: Eclipses at quiescence (2006 Apr 25)**
(note: in this plot *Magnitude* is the differential magnitude relative to the comparison star)

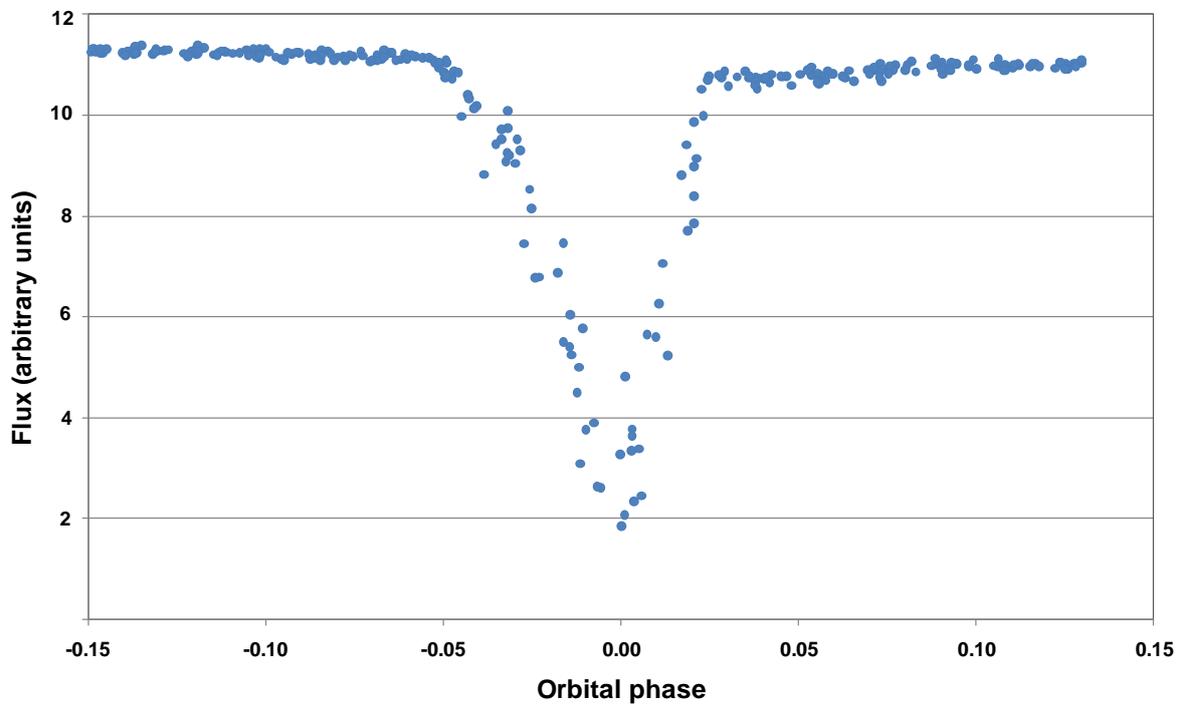

**Figure 10: Average quiescence eclipse**